\documentclass[12pt]{article}

\usepackage{times}
\usepackage[version=4]{mhchem} 
\usepackage{lineno} 
\usepackage{outlines}
\usepackage{amsmath}
\usepackage{graphicx}
\usepackage{hyperref}
\usepackage[super,sort&compress]{natbib}
\usepackage{subfiles}

\newenvironment{sciabstract}{%
\begin{quote} \bf}
{\end{quote}}

\title{Dissipative Kerr soliton photonic terahertz oscillator referenced to a molecule}

\author
{James Greenberg$^{\dag,\ast}$, Brendan M. Heffernan$^{\dag}$, Tomohiro Tetsumoto\\ and Antoine Rolland$^{\ast}$\\
\normalsize{IMRA America, Inc., Boulder Research Labs, Longmont, CO 80501, USA}\\
\normalsize{$^{\dag}$ These authors contributed equally.}\\
\normalsize{$^\ast$To whom correspondence should be addressed; E-mail: jgreenbe@imra.com, arolland@imra.com.}
}

\date{}

\begin{document}

\baselineskip24pt

\maketitle

\begin{sciabstract}
Controlling the coherence between light and matter has enabled the radiation of electromagnetic waves with spectral purity and stability that defines the Système International (SI) second. While transitions between hyperfine levels in atoms are accessible in the microwave and optical domains, faithfully transferring such stability to other frequency ranges of interest is not trivial. Such stability is specifically sought after for the terahertz domain to improve the resolution in very long baseline interferometry and molecular spectroscopy, and advance the technological development of high-speed, high data rate wireless communications. However, there is an evident lack of native frequency references in this spectral range, essential for the consistency of measurements and traceability. To mitigate the frequency drift encompassed in such waves, we experimentally demonstrate that using rotational spectroscopy of nitrous oxide (\ce{N2O}) can lead to linewidth reduction up to a thousandfold. A pair of diode lasers, optically injected with a low-noise, chip-based dissipative Kerr soliton, were incident upon a uni-travelling-carrier photodiode. We frequency-locked the emitted terahertz wave to the center of a rotational transition of \ce{N2O} through phase modulation spectroscopy. A terahertz wave with a 6~Hz linewidth was achieved (fractional frequency stability of $2 \times 10^{-11}$ at 1 second averaging  time) while circumventing the need of frequency multiplication or division of frequency standards.

\end{sciabstract}

\noindent
The scientific community has recently focused on efforts to fill the so-called terahertz gap~\cite{ozyuzer2007emission,sengupta2018terahertz,choi2022terahertz}. The gap refers to a lack of mature and convenient technologies for generating, manipulating, and detecting terahertz radiation. The terahertz band exists on the electromagnetic spectrum between microwaves and infrared light, both of which have well developed libraries of components for commercial and scientific use. Much of the motivation for bridging the gap is driven by a number of critical applications. For example, proposals for the sixth-generation wireless technology standard, 6G, call for carrier frequencies from 300\,GHz to many terahertz~\cite{Dang2020}. These larger frequencies are required to facilitate wider bandwidths for faster data rates and relieve congestion in currently deployed communication channels. Low noise and long-term stable THz oscillators are required to improve bit error rate and error vector magnitude for wireless data in order to realize this spectral band's full potential~\cite{katayama2016300,lee201980}. Another important application is very long baseline interferometry for radioastronomy where the coherence length of the local oscillator is needed to maintain constructive interference in the phased array antennae~\cite{yamada2006phase}.


 Generating terahertz radiation via photomixing two or more continuous wave (CW) optical frequencies on an ultra-fast diode offers many advantages~\cite{matsuura1997generation,Weiss:01,rolland2014narrow,Ishibashi2020}. This approach enables the use of widely available and high performance photonic technologies, including broadly tunable low noise lasers, optical frequency combs, fast modulators and photodiodes~\cite{Nagatsuma2016}. The technique can leverage advanced photonic integrated chip technology, like dissipative Kerr solitons (DKS) generated by micro ring resonators \cite{Kippenberg2018}, providing a path for low size, weight, power, and cost (SWAPc) terahertz devices. Importantly, photomixing currently provides the most spectrally pure THz waves. It was recently demonstrated that a pair of highly-coherent continuous wave lasers, divided down to the terahertz domain using DKS technology, led to the implementation of a compact 0.3~THz oscillator with a phase noise that out-preforms direct generation or multiplication of microwave references (for Fourier frequency above 100~Hz)~\cite{Tetsumoto2021}. 

Maintaining frequency stability in the long-term of such high performance terahertz oscillators remains a challenge. A terahertz wave generated via photomixing has a stability given by the quadratic sum of the optical waves' stability, which fractionally is seen from the terahertz spectral range as relatively mediocre. For example, beating two telecom lasers with $\sim$10~kHz linewidth (roughly corresponding to fractional stability in the order of $10^{-9}$) on a fast photodiode would lead to a $\sim$10~kHz linewidth as well, but at hundreds of GHz. At 0.3~THz, it corresponds to a fractional frequency stability of $10^{-6}$. While it is possible to improve this stability by phase locking the frequency difference of the two lasers to a microwave reference, this comes with costly and complex down-conversion schemes through electro-optic combs or sub-harmonic mixers~\cite{rolland2011non,zhang2019terahertz, tetsumoto2020300}. Additionally, the stability of the microwave reference can be corrupted in the process of converting it to the terahertz domain. To avoid the use of spectral purity transfer between disparate frequency ranges, we propose stabilizing a terahertz oscillator to a native terahertz reference through phase modulation spectroscopy of a gas~\cite{d2022terahertz}.

In the terahertz domain, many polar gas molecules (e.g., \ce{HCN}, \ce{OCS}, and \ce{N2O}) possess distinct transitions between quantized rotational states~\cite{townes2013microwave}. These levels are populated at room temperature by blackbody radiation. The dipole moments of such molecules give rise to absorption coefficients strong enough to perform absorption spectroscopy through a modest path length ($<1$\,m). Precise measurements of these molecular energy levels and lineshapes can reveal details of molecular structure as well as interactions between molecules and their environment. The most common practice is to probe a molecule with a terahertz source that is referenced to a microwave frequency standard~\cite{Wineland1979,mouret2009thz}. While many improvements to precision rotational spectroscopy techniques have been pioneered over the years~\cite{hsieh2016terahertz,bartalini2014frequency,Hindle2019,CAROCCI1996,Alighanbari2018,wang2018,Chou2020,ye1998ultrasensitive}, probing rotational transitions and using them to stabilize a DKS-based, photonic terahertz source is a fundamentally new approach.

Here, we constructed an oscillator consisting of a pair of diode lasers, separated in frequency by 301.442~GHz, which were simultaneously incident on a uni-travelling-carrier photodiode (UTC-PD). The resulting beat note was emitted by a horn antenna. To enable precision spectroscopy, the differential phase noise between the two diode lasers was minimized by optically-injecting two modes of a low noise DKS, respectively, in each diode. The frequency difference between the two laser diodes was locked to the center of a rotational transition via phase modulation spectroscopy of nitrous oxide (\ce{N2O}). This led to a 301.442~GHz oscillator with a sub-10~Hz linewidth and a fractional frequency stability of $2 \times 10^{-11}$ at 1 second averaging time, which corresponds to an unprecedented level of stability using a rotational transition in the terahertz domain.

\begin{figure*}[!ht]
    \centering
    \includegraphics[width=15cm]{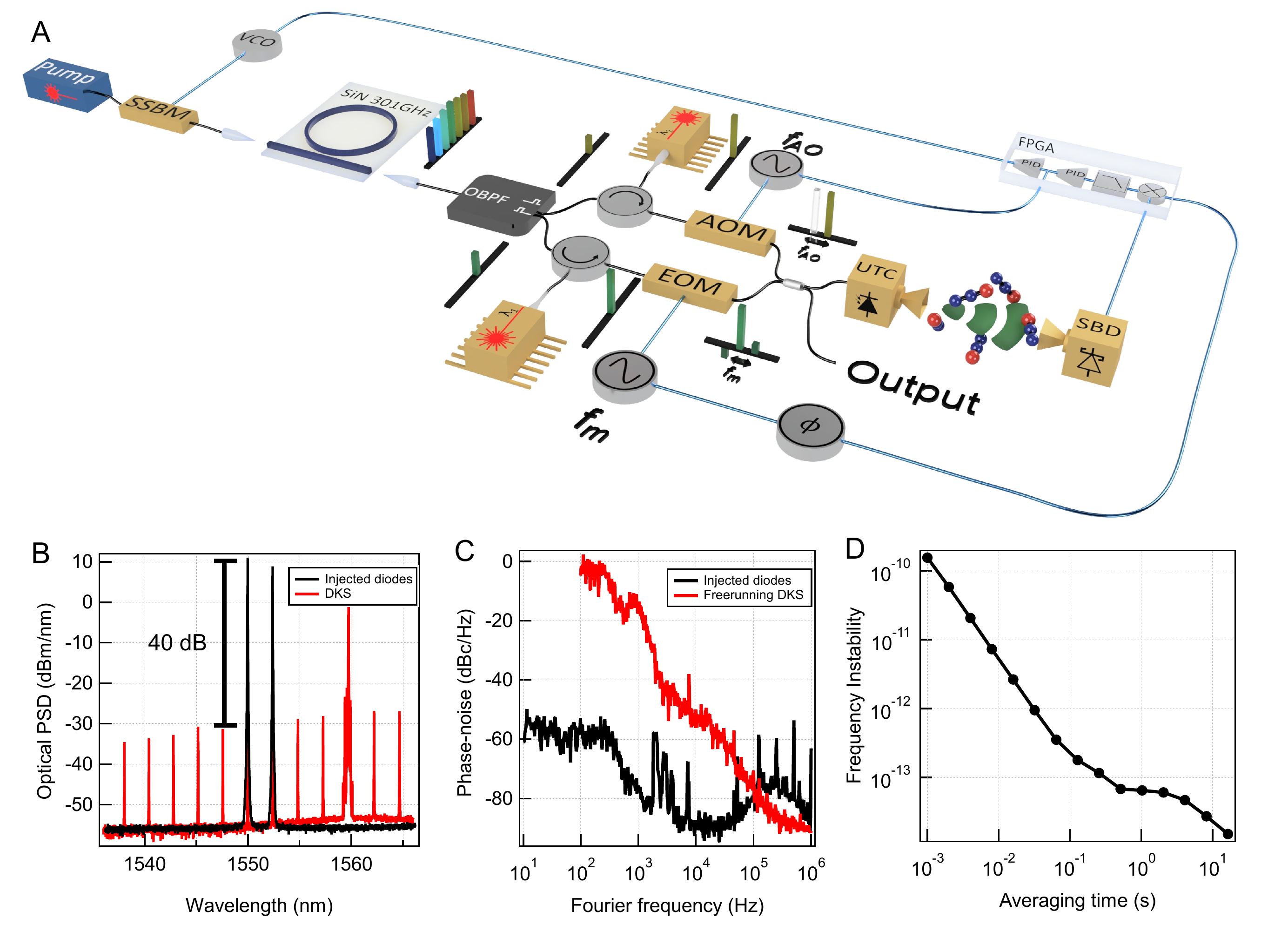}
    \caption{\textbf{A:} Concept of the terahertz oscillator based on two diode lasers pre-stabilized to a soliton microcomb through optical injection of two adjacent comb modes separated by the repetition rate frequency of 301.442 GHz. The differential phase noise between the two diode lasers is governed by the spectral purity of the soliton microcomb. One diode is phase modulated at a few MHz in order to perform phase modulation spectroscopy. The other diode is modulated through an acousto-optic modulator (AOM) to realize fine tuning of the frequency difference between the two diode lasers. The optical frequency difference is then converted into a terahertz wave through a uni-travelling carrier (UTC) photodiode emitting in a horn antenna. The wave propagates through gaseous nitrous oxide and rotational absorption is detected with a Schottky diode (also coupled to a horn antenna). Lock-in detection leads to the generation of a dispersion curve and the error signal is fed back to the voltage controlled oscillator driving the acousto-optic modulator AOM through a PID controller for fast control. The control signal is then fed through another PID filter and sent to the single side band modulator (SSBM) for slow control. \textbf{B:} Optical spectra of the output of the soliton microcomb and the combined diode lasers optically injected with the soliton. \textbf{C:} Additive phase noise induced by the spectral purity transfer of the microcomb to the two diode lasers.\textbf{D:} Residual instability induced by the spectral purity transfer of the microcomb to the two diode lasers.}
    \label{fig1}
\end{figure*}

Phase modulation spectroscopy relies on three core elements: a phase-modulated local oscillator, a gas cell, and a feedback mechanism that guarantees the local oscillator to be referenced to the gas. A conceptual overview of the system is shown in Fig. \ref{fig1}. In this experiment, the local oscillator was implemented with photonic technologies. DKS enable the generation of optical pulse trains with high repetition rates, usually on the order of a hundreds of GHz. Here, we generated a 301.442~GHz optical soliton with a microring on a silicon nitride (\ce{Si3N4}) chip (methods on how a soliton is generated are provided in the supplementary materials). While the phase noise properties and timing jitter of an optical soliton are attractive~\cite{lucas2020ultralow,tetsumoto2021effects,yao2022soliton}, using it directly for spectroscopy is not practical due to dispersive elements prior to photodetection, that inevitably destroy a spectroscopic signal. To entirely overcome dispersion engineering of an optical soliton, we transferred its spectral purity to a pair of diode lasers, separated by the same frequency of the optical soliton repetition rate. This was accomplished through optical injection locking~\cite{Stover1966, Liu2020a}, realized by first splitting the DKS comb into two paths. In this case, two neighboring comb lines at 1550~nm and 1552.42~nm were selected and each was sent to the input of the corresponding LD through a three-port fiber circulator. The optical spectrum of the output after filtering using a waveshaper is shown in Fig~\ref{fig1}(b). The injection-locked LD's show a 40~dB improvement in signal to noise ratio (SNR) compared to the DKS.

When attempting to transfer the spectral purity of an oscillator to another, a fundamental assessment is the residual noise measurement, i.e., verifying that the transfer process does not hinder the noise properties being transferred. We have measured the ability of two injection-locked LD's to reliably reproduce the phase noise properties of the DKS, as well as its stability (see the supplementary information for more experimental details). The phase noise of the DKS repetition rate and the residual phase noise of the injection process is shown on Fig.~\ref{fig1} (c) (black and red curve, respectively). For Fourier frequencies below 100~kHz, one can observe that the additive noise of injection-locking is well below the DKS repetition rate which indicates that the spectral purity transfer is successful. The bump above 100~kHz is due to the measurement noise floor (explanation given in the supplementary material). Additionally, we have evaluated the precision of the stability transfer in terms of Allan deviation. The residual fractional instability of injection-locking at 301.442~GHz reaches $8 \times 10^{-14}$ at 1 second averaging time. This corresponds to a linewidth transfer that is 25~mHz at 301.442~GHz while the free-running linewidth of the DKS repetition rate is larger than 1~kHz. Therefore, the spectral purity of the DKS repetition rate is faithfully transferred to a pair of LDs.

Prior to interfering the LDs on a UTC-PD, they were modulated by two optical modulators each serving a singular purpose. To perform phase modulation spectroscopy, one laser line was phase modulated with an electro-optic modulator (EOM) at the frequency $f_m$ to generate sidebands of the same frequency $f_m$ in the terahertz domain. The other laser line propagated through a deflecting acousto-optic modulator (AOM) shifting the laser frequency by $f_{AO}$ for fine frequency correction of the two lasers' frequency difference. This frequency difference, and thus the terahertz frequency, was locked to the molecular rotational transition via a PID loop from a FPGA-based signal processor. The FPGA performed lock-in detection as well as applied proportional and integrator gains to the control signal ultimately fed back to the VCO driving the AOM shifting the LD by $f_{AO}$. The phase difference between $f_{m}$ sent to the EOM and FPGA was tuned to guarantee the PID loop locked the terahertz wave to the peak of the absorption feature. Since the frequency range of the AOM is limited to a few MHz, we split the control signal into fast and slow components, and the slow component servoed the DKS pump laser frequency. This maintains pump-resonance detuning as the resonance of the SiN ring drifts due to environmental influences, stabilizing the repetition rate and guaranteeing robust operation. 

The terahertz power was monitored via the current controller powering the UTC-PD. The UTC-PD was biased with -1\,V and exposed to approximately 45\,mW of optical power. This pulled 9\,mA of photocurrent from the current controller, which corresponded to $\sim100\,\mu$W of radiated terahertz power. This power was stabilized via two methods. The first was the UTC-PD was temperature controlled ($\sim 295$\,K) using a thermo-electric cooler (TEC). Because the UTC-PD was in vacuum, temperature stabilization to $\pm 10$\,mK, the resolution limit of the temperature controller, was achieved. Second, the UTC-PD photocurrent was actively stabilized through a feedback loop between the current controller and the EDFA on the upper optical arm. Despite the EDFA being saturated, the current to the pump diode could be varied to make small adjustments to the total amplified optical power. These adjustments, along with temperature stabilization, held the UTC-PD photocurrent to within $\pm 10\mu$\,A of the 9\,mA setpoint.

\begin{figure*}[!ht]
    \centering
    \includegraphics[width=15cm]{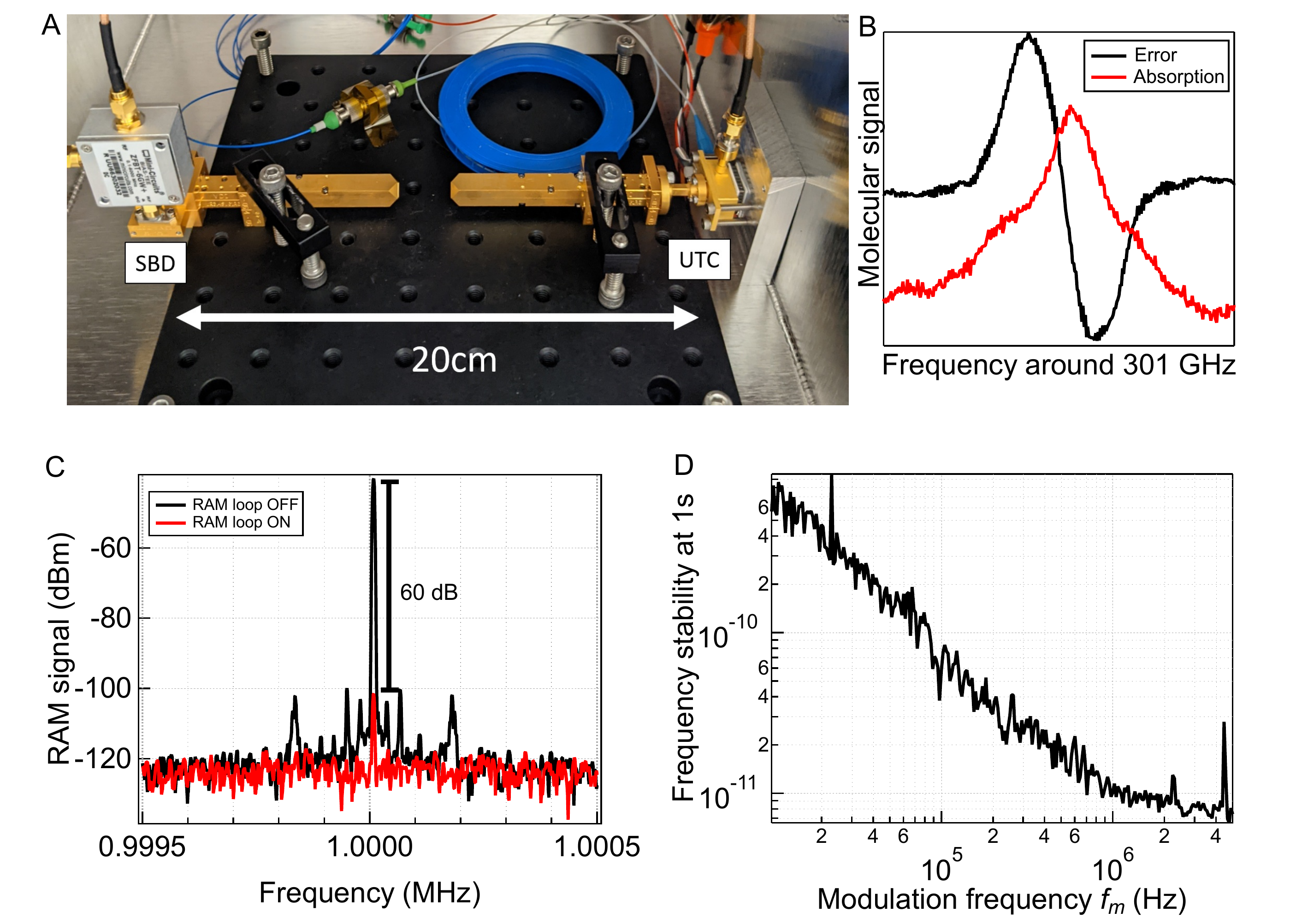}
    \caption{\textbf{\textbf{A:}} Photograph of the inside of the vacuum chamber where the uni-traveling carrier (UTC) photodiode, Schottky barrier diode (SBD), and their respective antennas are mounted. \textbf{B:} Measured molecular absorption line and its first-derivative dispersion curve at 301.442 GHz. \textbf{C:} RF spectra of the residual amplitude modulation (RAM) signal when not suppressed (black) and suppressed (red). \textbf{D:} Stability limit induced by the intermodulation effect derived from the free-running phase noise of the terahertz oscillator.}
    \label{fig2}
\end{figure*}

The terahertz radiation was generated at the UTC-PD and guided through a waveguide into a directional horn. A mirrored horn and waveguide directed the radiation onto a terahertz amplitude detector, which generated an electrical current proportional to the incident power. A bias-tee split the resulting electrical signal into DC and RF components which corresponded to the molecular absorption and error signals, respectively. A photograph of the absorption beam path is shown in Fig. \ref{fig2}(A). Both the UTC-PD source and Schottky detector were placed inside of a vacuum chamber with a base pressure of $\sim$20\,mTorr. Enclosing the terahertz antennas in the vacuum chamber is a preferred implementation of the gas cell because it avoids the use of a waveguide or cavity that lead to undesirable etalon effects~\cite{chevalier2019widely,Hindle2019,kim2019chip}. The vacuum chamber contained $\sim$50\,mTorr partial pressure of \ce{N2O}. A steady-state pressure of gas was maintained through the use of chamber pressure measurements at one second intervals, which fed back to a precision flow controller. This loop maintained the chamber pressure to $\pm 20\,\mu$Torr. The \ce{N2O} permeated the interior of the terahertz components, and thus utilized the entire path length between UTC-PD and Schottky diode for absorption ($\sim 20$\,cm). The terahertz frequency was tuned to the $J=11 \rightarrow J'=12$ transition resonance at 301.442\,GHz. By applying a ramp signal to the VCO driving the AOM, we could scan the terahertz frequency around the molecular resonance, and measure the absorption lineshape and its corresponding dispersion curve (see red and black curve, respectively, on Fig.~\ref{fig2}(B)). We measure a Doppler-limited linewidth of $\sim$1~MHz. The absorption signal was filtered with a low-pass filter at 1 kHz to remove residual DC amplifier noise from the data. The error signal, however, was left unfiltered to help estimate the signal-to-noise ratio (SNR). This error signal was used to qualitatively tune parameters in the experiment such as N2O gas pressure, UTC-PD photocurrent (a proxy for terahertz power), and its corresponding amplitude, and the phase difference between the signal sent to the diode laser and the lock-in detection. While in principle, the error signal SNR could limit the quality of the lock achieved by the PID loop, we found this did not limit lock performance in this experiment.

In phase modulation spectroscopy, minimizing the residual amplitude modulation (RAM) arising in the EOM is of utmost importance. RAM will indeed manifest itself in the error signal, adding an amplitude baseline shift in the spectroscopic feature and causing a frequency shift in the terahertz signal output. A fundamental issue is that this frequency shift will fluctuate over time, as it is related to two effects that depend on temperature variations: the orientation of the incident light polarization with respect to the principal axes of the birefringent electro-optic crystal and the etalon effects arising both within the crystal and along the light path in fibers after the phase modulator. To combat this undesirable effect, we have implemented a RAM cancellation scheme~\cite{Zhang2014} in which we, first, actively control the temparature of the electro-optic modulator with a $\pm 1$~mK precision and, second, we apply a DC electric field to the modulator through a bias-tee. The DC electric field is derived from an error signal generated with a photodiode that detects the residual amplitude modulation (see black spectrum Fig.~\ref{fig2}(C)). When this RF signal is demodulated with the same tone frequency $f_m$, we can measure the RAM offset fluctuation at DC. Nullifying this signal with a PID filter leads to a $\sim$ 60~dB cancellation of the RAM signal as can be seen on the red spectrum of Fig.~\ref{fig2}(C).

Another important limit in phase modulation spectroscopy is caused by the intermodulation effect~\cite{Audoin1991}. The phase noise of the terahertz oscillator, at Fourier frequencies equal to even multiples of the modulation frequency $f_m$, is transferred into the error signal. Then an additional perturbation of the terahertz oscillator arises, which sets a limit to its achievable frequency stability. Once the phase noise of the oscillator is known, it is possible to derive the stability limit caused by the intermodulation effect. In the present case, the stability limit in terms of Allan deviation at 1~second averaging time as a function of modulation frequency $f_m$ is shown on Fig.~\ref{fig2}(D). For a modulation frequency $f_m$ higher than 1~MHz, the stability limit induced by the intermodulation effect is lower than $1 \times 10^{-11} / \sqrt\tau$. Of course, shot noise and thermal noise in the UTC-PD and the Schottky diode will also contribute to the signal-to-noise ratio of the error signal that will also limit the stability.

\begin{figure*}[!ht]
    \centering
    \includegraphics[width=10cm]{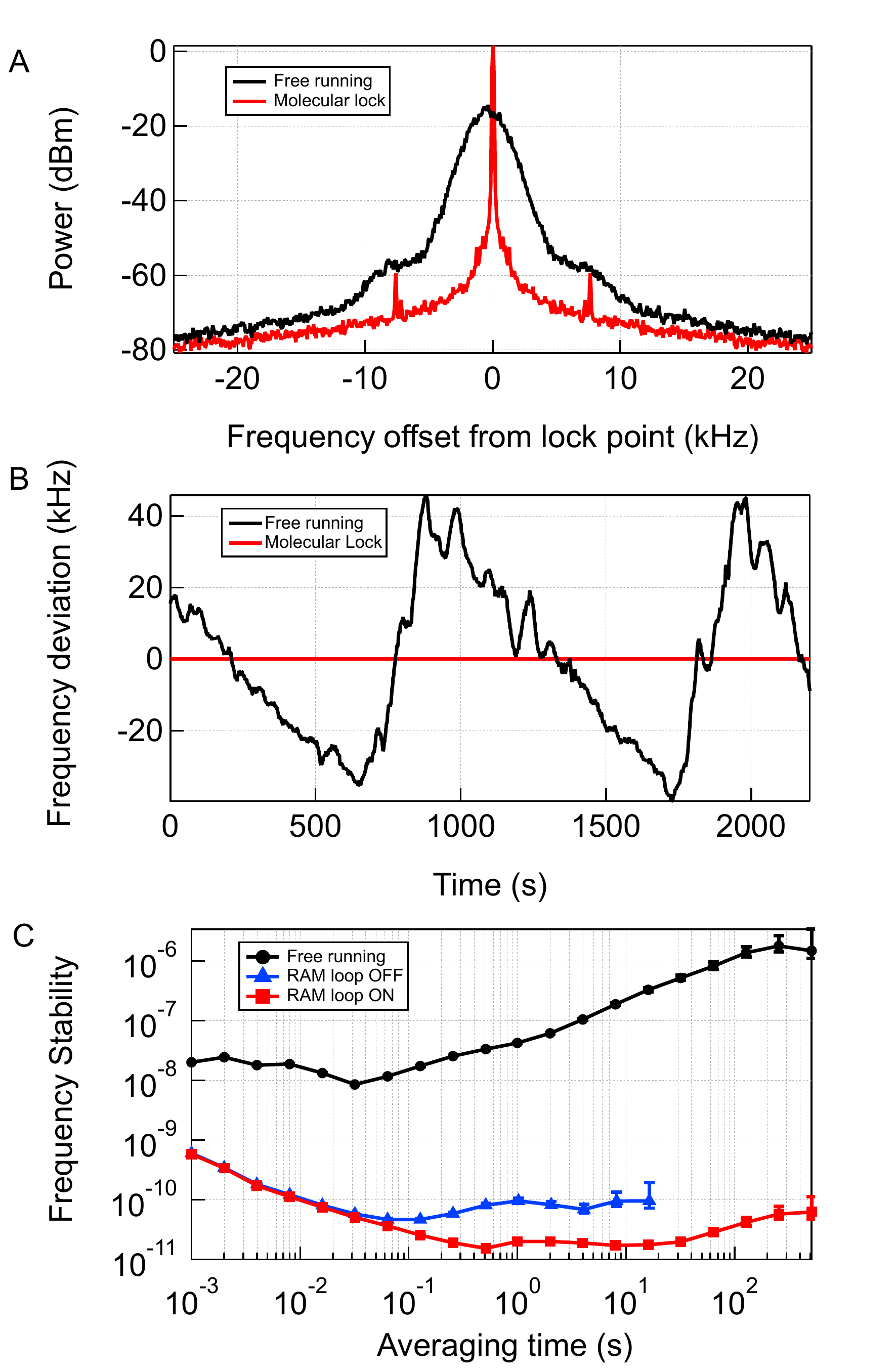}
    \caption{\textbf{A:} RF spectra of the terahertz oscillator in free-running (black spectrum) and referenced to \ce{N2O}. \textbf{B:} Terahertz oscillator absolute frequency over time when free-running (black) and locked to \ce{N2O} (red). \textbf{C:} Fractional frequency instability in terms of modified Allan deviation when in free-running (black), when locked to \ce{N2O} with RAM cancellation OFF (blue) and when locked to \ce{N2O} with RAM cancellation ON (red).}
    \label{fig3}
\end{figure*}

To assess the performance of the terahertz oscillator locked to a \ce{N2O} rotational line, we need to conduct a so-called out-of-loop measurement, in which we compare the stability of the terahertz wave to an identical or better reference. Because the frequency of the terahertz radiation was equal to the frequency difference of the LDs, we measured the frequency difference via optical down-conversion~\cite{rolland2011non}. The light, split before the UTC photodiode, was amplified using an EDFA and sent through an electro-optic (EO) comb. Both laser lines passed through three cascaded EO modulators, driven by a synthesizer referenced to a 10~MHz stable rubidium (Rb) reference. The EO-comb generated more than 15 sidebands (comb-modes) with a spacing of $f_{synth}\sim$10\,GHz between modes. The sidebands spanned the gap between the two laser-frequencies and produced a set of overlapping modes that were isolated using an optical band-pass filter (OBPF). The isolated lines were photodetected and produced a beat-note with a frequency that was low enough ($f_{beat}<$1\,GHz) to be measured by a conventional spectrum analyzer. This frequency was directly related to the terahertz frequency by \begin{equation}
    f_{THz} = \Delta f_{diode} = 2nf_{synth} \pm f_{beat}
\end{equation}
where $n$ was the n-th order comb mode. The sign of the $f_{beat}$ term was determined experimentally by changing $f_{synth}$ a small amount and observing the resulting shift in $f_{beat}$. The stability of 2n$f_{synth}$ is directly derived from the stability of the Rb reference making the measurement noise floor limited by the stability of the Rb reference. By observing the RF spectrum of the down-converted terahertz wave $f_{beat}$, we can observe the impact of spectroscopic feedback on the signal linewidth (see Fig.~\ref{fig3}(A)). The free running local oscillator (black spectrum) exhibits a linewidth that is larger than a few kHz. In closed loop operation the oscillator linewidth is dramatically reduced (red spectrum). 

However, to fully characterize the impact of referencing the terahertz oscillator to the molecule rotational transition at longer timescales, we count its absolute frequency against the Rb reference at 10 MHz. The beat-note frequency $f_{beat}$ was sent to a zero-deadtime frequency counter with a 1~ms gate time, also referenced to the Rb clock. The frequency counter tracked the absolute frequency of the terahertz oscillator over time. Fig.~\ref{fig3}(B) illustrates the striking long-term stability improvement. Fluctuations of the free-running terahertz oscillator swings by almost 100~kHz in the course of 30~min (black trace), while the locked operation mitigates those fluctuations below 100~Hz (red trace). By acquiring the absolute frequency of the terahertz oscillator, we can compute the Allan deviation (most standard criteria to characterize the stability of an oscillator). Due to the relatively low signal-to-noise ratio of $f_{beat}$ (white phase noise limited), we chose to compute the modified Allan deviation in which the white phase noise floor is filtered allowing for a more accurate estimation of long-term stability. On Fig.~\ref{fig3}(C), we plot the stability of the terahertz oscillator in three configurations: free-running (black curve), locked to the molecule without RAM cancellation (blue curve) and locked to the molecule with RAM cancellation (red curve). At 1 second averaging time, the free-running terahertz oscillator stability is $6 \times 10^{-8}$, suggesting an 18~kHz linewidth, and it drifts away to $1 \times 10^{-6}$ at about 100 seconds averaging time. When the terahertz oscillator is referenced to the molecule without compensating the effect of RAM, the stability at 1 second averaging time is reduced to $1 \times 10^{-10}$. When we compensate for the RAM, the stability is then pushed to $2 \times 10^{-11}$ (linewidth lower than 10~Hz). We chose the modulation frequency $f_m$ to be 1~MHz (similar to the rotational transition linewidth). The intermodulation limit given the free running phase noise of the terahertz oscillator is indeed $1 \times 10^{-11} / \sqrt\tau$. At 100 seconds averaging time, the frequency drift is clearly damped compared to free-running operation with a stability at $5 \times 10^{-11}$, which this time corresponds to a drift mitigation by a factor of 20,000. These stability measurements highlight a few facts that we anticipated. Referencing a terahertz oscillator to a molecular rotational transition is an effective method to dramatically reduce the linewidth, here by more than a thousandfold. By compensating the RAM, we can get into the intermodulation limit at $<$1 second averaging time caused by the phase noise of the terahertz oscillator.

This demonstration only aimed at reducing the linewidth of a terahertz oscillator using direct spectroscopy of a molecule in the terahertz domain. However, understanding the mechanism that prevents the oscillator to be limited by the white frequency noise behavior ($1/\sqrt\tau$), inherent in referencing an oscillator to an atom or a molecule, is essential in order to realize the yet elusive goal of a terahertz frequency standard. At longer averaging times, the oscillator's drift is ascribed to thermal fluctuations in the laboratory. Lacking a detailed understanding of the mechanism involved, we could improve the long-term stability by using a molecule with a stronger molecular absorption feature. The fractional effect of spectroscopic baseline fluctuations will ultimately be smaller when the amplitude of the error signal increases. There are two main contributions to the strength of the absorption feature. The first is the electric dipole moment of the molecule. The second is the occupation of the quantum state being measured at room temperature. Carbonyl sulfide (\ce{OCS}) would be a better choice of molecule than \ce{N2O} in both regards. For the same path length, we would expect a larger absorption and corresponding error signal amplitude for \ce{OCS}, leading to a reduction in sensitivity to temperature induced frequency shifts. Thus we expect to obtain better long-term performance and lower fractional frequency stability limit by switching to \ce{OCS}. Improving the short-term stability will rely on better free-running phase noise of the terahertz oscillator. This can be made possible by locking the DKS to a low noise optical reference~\cite{Tetsumoto2021} or by operating the DKS in its quiet regime~\cite{liu2020photonic,tetsumoto2021effects} and support stability in the order of $\sim 1 \times 10^{-12}/\sqrt\tau$. At the expense of compactness, the use of a terahertz source based on an optical fiber cavity~\cite{amin2022exceeding}, could lead to intermodulation limit of $ \sim 1 \times 10^{-13}/\sqrt\tau$.

Despite the current limitations in our oscillator performance, the mean of the frequency data presented in Fig. \ref{fig3}(C) is 301,442,731,700\,Hz and has a standard deviation of only 266\,Hz. This statistical uncertainty corresponds to the most precise measurement of the J=11 rotational transition frequency ever published. Precision of this level in other room temperature gas samples is not unprecedented, but this experiment lacks the influence of any kind of cavity or etalon that most precision microwave spectroscopy must contend with \cite{Hindle2019,CAROCCI1996}. While a full analysis of systematic uncertainties are outside the scope of this work, we believe the advantage of precision without a cavity will lead to breakthroughs in accuracy as well. This is supported by good agreement between our measurement, and thus far the most accurate measurement published: of $301,442,710,000 \pm 50,000$\,Hz \cite{Kropnov1974} (via NIST molecular spectral database). Since our experiment can be easily generalized to many molecules and many rotational levels, it may provide a powerful new technique for precision molecular rotational spectroscopy.

The prospect of developing a terahertz oscillator into a standard frequency reference are initially encouraging. Setting new accuracy limits for precision molecular rotational spectroscopy is the next critical step to be taken. The presented short-term stability already rivals that of the commercially available Rb-clock reference utilized in this experiment. With the potential improvements in both short and long-term performance discussed in this section, photonic terahertz oscillators disciplined by molecular rotations may be serious candidates for future frequency references. At a minimum, a similar apparatus can be used as a frequency reference for terahertz applications in which a native terahertz reference is necessary or access to another SI standard is impractical.

The implications of this work are multiple. The terahertz local oscillator can be implemented in an ultra-compact fashion as it was designed and built with only low-cost photonics components at telecom wavelengths. The oscillator's low noise properties are owed to the use of a DKS generated from a \ce{Si3N4} chip that can, nowadays, be battery-operated~\cite{stern2018battery}. Additionally, there is no need to self-reference the DKS, which significantly simplify the execution of the oscillator. Instead, referencing the terahertz oscillator to the rotational spectrum of molecules at room temperature is a very attractive solution. Apart from the fact that sub-10Hz terahertz linewidth is achievable, molecules offer rich structures that can fulfill different tasks. It is realistic to stabilize a terahertz oscillator anywhere in the rotational spectrum which exists a rotational transition. In the case of \ce{N2O}, every $\sim$ 25~GHz from 0.2 to 1.2~THz are practically available, and could obtain a form of channelization of local oscillators that is essential for wireless communication. Finally, this work also paves the way towards new spectroscopic techniques that could lead to compact spectrometers for gas sensing applications.

\section*{Acknowledgment}
We thank Martin Fermann and Mark Yeo for their contributions in the early stage of this work. The authors are thankful to Rubab Amin for assembling the electro-optic comb used to down-convert to baseband the terahertz oscillator. We thank Tara Fortier from the National Institute of Standards and Technology for her help and support.

\section*{Authors contribution}
JG and AR designed the architecture of the setup. JG built the science chamber and all the spectroscopic aspects of the setup. BH built the terahertz local oscillator. JG and BH operated the experimental setup and acquired the data. JG, BH and AR analyzed the data. TT designed the silicon nitride chip. AR initiated and supervised the project.

\bibliography{ms}
\bibliographystyle{unsrtnat}

\newpage

\end{document}